# Astroinformatics: Statistically Optimal Approximations of Near-Extremal Parts with Application to Variable Stars

Andronov, Ivan L.[*1], Andrych, Kateryna D.[†1], Chinarova, Lidia L.[‡2,1], and Tvardovskyi, Dmytro E.[§1,3]

[1]Department of Mathematics, Physics and Astronomy, Odessa National Maritime University, Odessa, Ukraine
[2]Astronomical Observatory, Odessa I.I.Mechnikov National University, Odessa, Ukraine
[3]Department of Theoretical Physics and Astronomy, Odessa I.I.Mechnikov National University, Odessa, Ukraine

## Abstract

The software MAVKA is described, which was elaborated for statistically optimal determination of the characteristics of the extrema of 1000+ variable stars of different types, mainly eclipsing and pulsating.

The approximations are phenomenological, but not physical. As often, the discovery of a new variable star is made on time series of a single-filter (single-channel) data, and there is no possibility to determine parameters needed for physical modelling (e.g. temperature, radial velocities, mass ratio of binaries).

Besides classical polynomial approximation "AP" (we limited the degree of the polynomial from 2 to 9), there are realized symmetrical approximations (symmetrical polynomials "SP", "wall-supported" horizontal line "WSL" and parabola "WSP", restricted polynomials of non-integer order based on approximations of the functions proposed by Andronov (2012) and Mikulasek (2015) and generally asymetric functions (asymptotic parabola "AP", parabolic spline "PS", generalized hyperbolic secant function "SECH" and "log-normal-like" "BSK").

This software is a successor of the "Observation Obscurer" with some features for the variable star research, including a block for "running parabola" "RP" scalegram and approximation. Whereas the RP is oriented on approximation of the complete data set.

MAVKA is pointed to parts of the light curve close to extrema (including total eclipses and transits of stars and exoplanets). The functions for wider intervals, covering the eclipse totally, were discussed in 2017Ap.....60...57A . Global and local approximations are reviewed in 2020kdbd.book..191A .

The software is available at http://uavso.org.ua/mavka and https://katerynaandrych.wixsite.com/mavka.

We have analyzed the data from own observations, as well as from monitoring obtained at ground-based and space (currently, mainly, TESS) observatories. It may be used for signals of any nature.

## 1. Introduction

The variety of types of variability, as well as gaps in the astronomical observations, need a wide net of methods for analysis, which would allow statistically optimal determination of parameters using adequate approximations of the data. Many hundreds of monographs were published, the majority of them devoted to continuous signals or, at least, to signals, which are equidistantly spaced in time: $t_k = t_L + (k - L)\delta$. Here, $k, L = 1..n$, $n$ is the number of observations, and $\delta-$ is the interval between each pair of the subsequent observations (sometimes called the "Time resolution").

[*]tt_ari@ukr.net, Corresponding author
[†]katyaandrich@gmail.com
[‡]llchinarova@gmail.com
[§]dtvardovskyi@ukr.net

In this paper, we describe the software MAVKA ("Multi-Analysis of Variability by Kateryna Andrych") elaborated mainly for determination of characteristics of extrema, primarily, Times of Maxima/Minima (current abbreviation "ToM", also known as "minima timings").

The software is a successor (add-on) of the computer program OO ("Observational Obscurer") and it's few realizations in different computer languages/operational systems (e.g. Andronov (2001)). The "running parabola" ("RP") scalegram and approximation of the complete data set (Andronov (1997)) are available.

The logo of MAVKA software is the letter "M" with 3 points showing a triple nature (third body at a distant orbit around an eclipsing binary) of the majority of objects we studied.

## 2. Why Phenomenological Modelling?

The aim of physical modelling is to determine physical parameters (e.g. masses, radii of stars) from the observations. However, till now, only 305 binary systems have been studies completely, with determination of masses, parameters etc. caleb.eastern.edu, while the number of catalogued eclipsing systems is 8955 (192 combinations of types) in the current GCVS (General Catalogue of Variable Stars, Samus et al. (2017), www.sai.msu.su/gcvs/). In the Variable Stars Index (VSX, aavso.org/vsx), there are currently 2105479 objects. So, only one system among 7000 is studied sufficiently. For the rest 99.986%, the only available methods of study are phenomenological modelling.

There are dozens of types of variability and hundreds of combinations, so there are some basic principles instead of a single method, There are numerous reviews on step-by-step improvements and extensions of the methods (Andronov (1994), Andronov (2003), Andronov (2005), Andronov (2020), Andronov et al. (2020a)).

The main idea was to make an (nearly) "all in one" software, which realizes main methods proposed in our group, as well as most important methods proposed by others.

## 3. Algorithms used in MAVKA

The algorithms with formulae were described by Andrych & Andronov (2019).

For an illustration, we have used the small part of observations (102 out of 629 in the filter V) of the pulsating variable RR Lyr made by Horace Smith (AAVSO code SHA). The approximations are shown in Fig. 2.

The methods may be grouped as follows:

- Generally asymmetric extrema, which are common in pulsating variables except RRc and DSct. The commonly used functions are polynomials of statistically optimal order (cf. Chinarova & Andronov (2000). Andrych et al. (2015) compared these approximations with the "Asymptotic Parabola" (AP) fits (Marsakova & Andronov (2015)). The "spline-like" extension of the AP function, is the parabolic spline (PS) (Andrych et al. (2020a)). For longer intervals of data, which are extended to opposite nearby extrema (e.g. to minima before and after the maximum), there are analytical functions, e.g. BSK (B´odi et al. (2016)) and SECH (asymmetrical hyperbolic secant, Andronov (2005)). SECH is more stable, makes asymptotic lines in a logarithmic scale (Fig.3). This allows determination of characteristic times of rise and fall of luminosity, if using the intensities instead of stellar magnitudes. BSK works exclusively with asymmetrical extrema. For positive parameters, the larger absolute slope at the preceding branch is needed. The function often has problems with convergence of the parameters to finite values. However, it was included to MAVKA for realizing existing methods, even if we do not recommend to use them.

- Generally symmetric extrema. Such approximations are common for the eclipses in the binary systems. For these, we implement symmetric polynomials with even powers of deviation $(t-\bar{e})^{2j}$ of argument from the time of extremum $t_e$. Also, we used the abbreviated Taylor series for the functions proposed by Mikulášek (2015) and Andronov (2012).

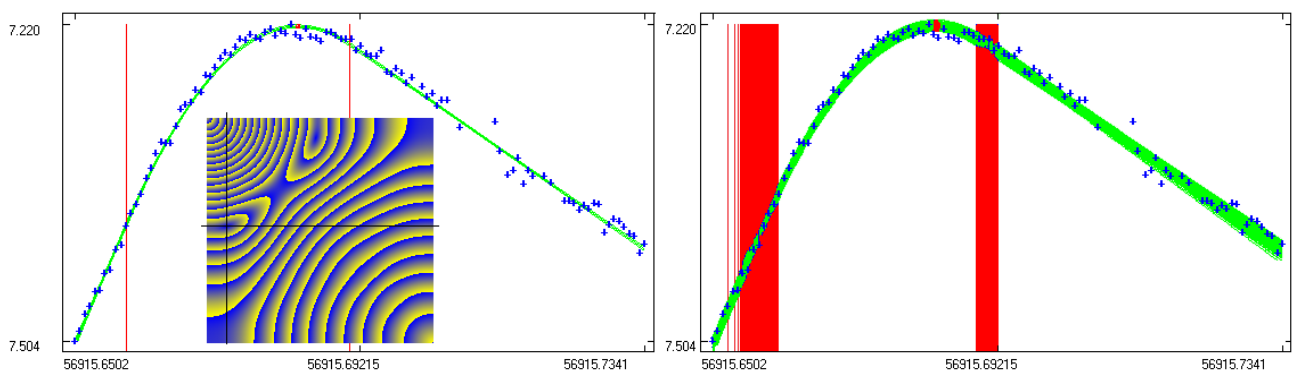

Figure 1. "Asymptotic parabola" approximation of the near-eclipse part of observations of RR Lyr using the program MAVKA. *Left:* the screenshot with the data (blue), approximation and $\pm 1\sigma$ "error corridor". Vertical red lines show the borders of intervals with lines connected with a parabola. The yellow-blue 2D dependence of the test function on positions of the borders of the inner interval, the black lines show position of the minimum of the test functions. *Right:* The approximations of the 2000 pseudo-data generated using the bootstrap algorithm. This sample data show much more stable bootstrap approximations than that discussed by Andrych et al. (2020a).

Andrych et al. (2017) introduced the "Wall-Supported Polynomial" (WSP) algoritms for statistically optimal modeling of flat eclipses and exoplanet transitions.

Obviously, these approximations are bad for asymmetric extrema, as one may see e.g. at the right part of Fig. 2.

- The moments of crossings of some fixed value $m_0$. In this case, it is recommended to use inverse approximation $t(m)$ and it's value $t(m_0)$ (e.g. Andronov & Andrych (2014)).

Andrych et al. (2020b) and Tvardovskyi et al. (2020a) investigated the approximation stability for various methods implemented in MAVKA.

## 4. Applications to Concrete Stars

### 4.1. Eclipsing Binary Systems

During our studies of (O-C) diagrams, the initial data were taken either from own CCD photometrical observations, or the ground-based or space surveys. The original photometric observations were distributed into intervals near extremum and then analyzed with MAVKA.

Tvardovskyi et al. (2017) studied effects of the Mass Transfer and Presence of the Third Components in Close Binary Stellar Systems. Tvardovskyi et al. (2018) detected period variations and possible third components in the eclipsing binaries AH Tauri and ZZ Cassiopeiae. Tvardovskyi et al. (2020b) elaborated the code to model third components with elliptical orbits in the eclipsing binaries. Tvardovskyi (2020a) determined parameters of AB Cas, AF Gem, AR Boo, BF Vir and CL Aur. The catalogue of moments of Minima of 25 Eclipsing Binaries was computed by Tvardovskyi (2020b).

MAVKA is also used as a method complementary to the approximation of the complete light curves using the NAV ("New Algol Variable") algorithm (Andronov (2012), Tkachenko et al. (2016)). An example of such combined study is two eclipsing binaries V454 Dra and V455 Dra in the field of cataclysmic variable DO Draconis is shown by Kim et al. (2020).

A review on highlights of studies of interacting binary stars with instability of the light curves using these (and other) methods is presented by Andronov et al. (2017).

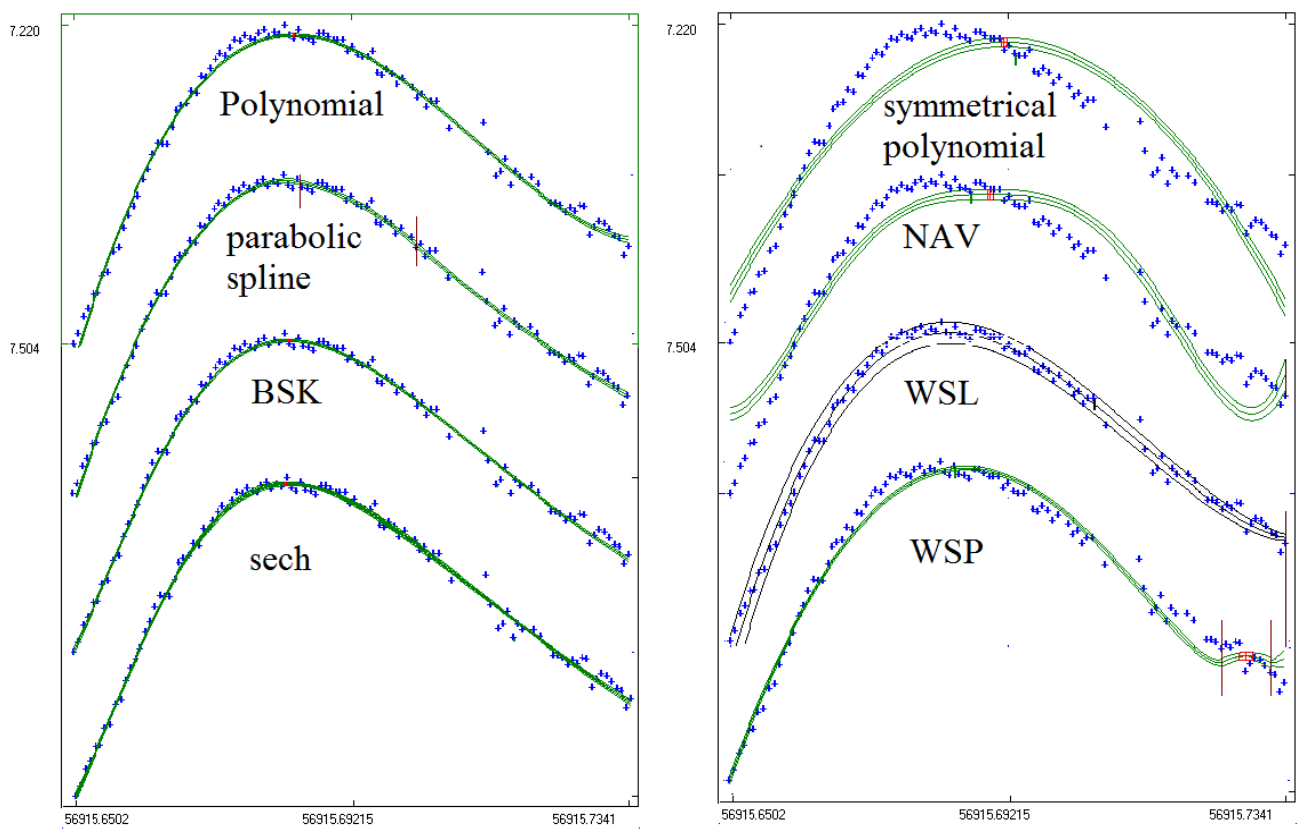

Figure 2. *Left:* approximations, which are generally oriented on asymmetric extrema, and show adequate quality of the fit. *Right:* The "bad" approximations of the asymmetric data using symmetric functions.

## 4.2. Intermediate Polars

Intermediate polars show a complicated behaviour because of at least two-period variations.One period is due to the orbital motion of a cataclysmic binary, and another period is due to "spin" (rotation of the magnetic white dwarf) and thus changing orientation of the accretion columns. In many cases, two accretion columns are seen alternatelySo there are significant waves not only with the main (spin) period, but also the wave with a double frequency.

Such studies were made for nearly a dozen of intermediate polars. Two-Color CCD Photometry of the Intermediate Polar 1RXS J180340. 0 +401214 showed a remarkably stable period (Andronov et al. (2011)).

The variability of the spin period of the white dwarf in the intermediate polar V405 Aur is present, however, its type currently does not allow to decide, whether these variations are due to a light-time effect caused by a low-mass third body, or to the precession of the magnetic white dwarf surrounded by a (warped) accretion disk (Breus et al. (2013)).

A fast spin period decrease was detected in the intermediate polars V2306 Cygni (Breus et al. (2019)) and period variations of FO Aqr were studied (Breus et al. (2012)).

However, the phases of individual cycles of spin variability of MU Cam show a wave with an orbital phase (Kim et al. (2005)), which apparently may be interpreted as the orbital sideband of the spin period (Parimucha et al. (2020)).

## 4.3. Pulsating Stars

For the pulsating stars, we make a few-component analysis including period search and determination of the statistically optimal degree of the trigonometrical polynomial, phase plane analysis of the photometrical variations (Kudashkina & Andronov (2017b)).

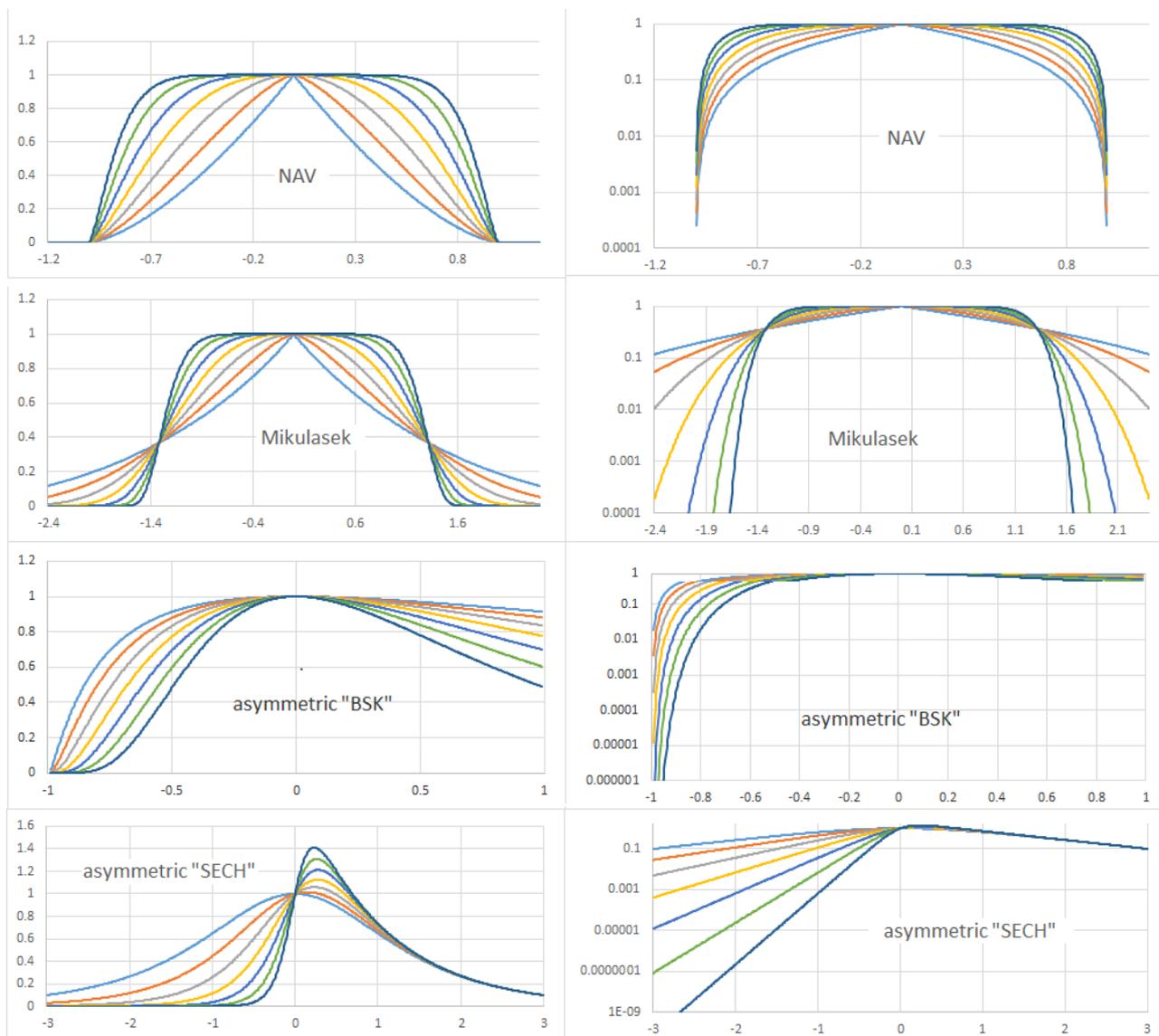

Figure 3. Some of functions implemented in MAVKA in linear *(left)* and logarithmic *right* scales.

Andronov et al. (2020b) made multi-algorithm analysis of the semi-regular variable DY Per, the prototype of the class of cool RCrB variables.

Impact of Pulsation Activity on the Light Curves of Symbiotic Variables was studied by Marsakova et al. (2015).

Wavelet Analysis of Semi-Regular Variables was made by Chinarova (2010) and Kudashkina & Andronov (2010).

A recent review on semi-regular variable stars is presented by Kudashkina (2019).

### Acknowledgements

We thank Dr. Areg Mickaelian for fruitful discussions and the AFOEV (cdsarc.u-strasbg.fr/afoev/) and AAVSO (aavso.org) for numerous observations of variable stars. Special thanks to Horace Smith (AAVSO code SHA) for his photometrical observations of RR Lyr, some of which were used for the illustration of our methods.

Andronov I. L., 2001, Odessa Astronomical Publications, 14, 255

Andronov I. L., 2003, ASP Conference Series, 292, 391

Andronov I. L., 2005, ASP Conference Series, 335, 37

Andronov I. L., 2012, Astrophysics, 55, 536

Andronov I. L., 2020, Knowledge Discovery in Big Data from Astronomy and Earth Observation, 1st Edition. Edited by Petr Skoda and Fathalrahman Adam. ISBN: 978-0-128-19154-5. Elsevier, pp 191–224

Andronov I. L., Andrych K. D., 2014, Odessa Astronomical Publications, 27, 38

Andronov I. L., Kim Y., Yoon J.-N., et al 2011, Journal of The Korean Astronomical Society, 44, 89

Andronov I. L., et al., 2017, ASP Conference Series, 511, 43

Andronov I. L., Breus V. V., Kudashkina L. S., 2020a, "Development Of Scientific Schools Of Odessa National Maritime University": Collective monograph. Riga, Baltija Publishing, 0, 3

Andronov I. L., Andrych K. D., Chinarova L. L., 2020b, Annales Astronomiae Novae, 1, 179

Andrych K. D., Andronov I. L., 2019, Open European Journal on Variable Stars, 197, 65

Andrych K., Andronov I. L. Chinarova L. L., Marsakova V. I., 2015, Odessa Astronomical Publications, 28, 158

Andrych K. D., Andronov I. L., Chinarova L. L., 2017, Odessa Astronomical Publications, 30, 57

Andrych K. D., Andronov I. L., Chinarova L. L., 2020a, Journal of Physical Studies, 24, 1902

Andrych K. D., Tvardovskyi D. E., Chinarova L. L., Andronov I. L., 2020b, Contributions of the Astronomical Observatory Skalnate Pleso, 50, 557

Breus V., Andronov I., Hegedus T., Dubovsky P., Kudzej I., 2012, Advances in Astronomy and Space Physics, 2, 9

Breus V., et al., 2013, Journal of Physical Studies, 17, 3902

Breus V., Petrik K., Zola S., 2019, Monthly Notices of the Royal Astronomical Society, 488, 4526

Bódi A., Szatmáry K., Kiss L. L., 2016, Astronomy & Astrophysics, 596, id.A24, 8 pp.

Chinarova L. L., 2010, Odessa Astronomical Publications, 23, 25

Chinarova L. L., Andronov I. L., 2000, Odessa Astronomical Publications, 13, 116

Kim Y.-G., Andronov I., Park S.-S., Chinarova L., Baklanov A., Jeon Y.-B., 2005, Journal of Astronomy and Space Sciences, 22, 197

Kim Y., Andronov I. L., Andrych K. D., Yoon J.-N., Han K., Chinarova L. L., 2020, Journal of Korean Astronomical Society, 53, 43

Kudashkina L., 2019, Astrophysics, 62, 556

Kudashkina L. S., 2020, Annales Astronomiae Novae, 1, 199

Kudashkina L., Andronov I., 2010, Odessa Astronomical Publications, 23, 67

Kudashkina L., Andronov I., 2017a, Czestochowski Kalendarz Astronomiczny, 14, 283

Kudashkina L., Andronov I., 2017b, Odessa Astronomical Publications, 30, 93

Marsakova V. I., Andronov I. L., 2015, Odessa Astronomical Publications, 28, 158

Marsakova V. I., Andronov I. L., Chinarova L. L., Chyzhyk M. S., Andrych K. D., 2015, Czestochowski Kalendarz Astronomiczny, 12, 269

Mikulášek Z., 2015, Astronomy & Astrophysics, 584, id.A8, 13 pp.

Parimucha S., Dubovsky P., Kudzej I., Breus V., Petrik K., 2020, Contributions of the Astronomical Observatory Skalnat´e Pleso, 50, 618

Samus N., Kazarovets E., Durlevich O., Kireeva N., E.N. P., 2017, Astronomy Reports, 61, 80

Tkachenko M., Andronov I. L., Chinarova L., 2016, Journal of Physical Studies, 20, 4902

Tvardovskyi D., 2020a, Advances in Astronomy and Space Physics, 10, 55

Tvardovskyi D. E., 2020b, Open European Journal on Variable stars, 204, 1

Tvardovskyi D. E., Marsakova V. I., Andronov I. L., 2017, Odessa Astronomical Publications, 30, 135

Tvardovskyi D. E., Marsakova V. I., Andronov I. L., Shakun L. S., 2018, Odessa Astronomical Publications, 31, 103

Tvardovskyi D. E., Andronov I. L., Andrych K. D., Chinarova L. L., 2020a, in Proceedings of the conference "Stars and their Variability Observed from Space". pp 381–382

Tvardovskyi D. E., Marsakova V. I., Andronov I. L., 2020b, Journal of Physical Studies, 24, 3904